\documentclass[aps,prl,twocolumn,superscriptaddress,amsmath,amssymb,
]{revtex4-1}

\usepackage{graphicx}
\usepackage{dcolumn}
\usepackage{bm}
\usepackage[linewidth=1pt]{mdframed}

\usepackage{siunitx} 
\usepackage{hyperref}

\hypersetup{
	colorlinks=true,
	linkcolor=blue,
	citecolor=blue,
	urlcolor=blue
}

\begin{document}


\title{Linear Ultrastrong Optomechanical Interaction}

\author{Kahan Dare}
 \affiliation{University of Vienna,  Faculty of Physics, Vienna Center for Quantum Science and Technology, A-1090 Vienna, Austria}
 \affiliation{Institute for Quantum Optics and Quantum Information (IQOQI) Vienna, Austrian Academy of Sciences, A-1090 Vienna, Austria.}
\author{Jannek J. Hansen}
 \affiliation{University of Vienna,  Faculty of Physics, Vienna Center for Quantum Science and Technology, A-1090 Vienna, Austria}
 \author{Iurie Coroli}
 \affiliation{University of Vienna,  Faculty of Physics, Vienna Center for Quantum Science and Technology, A-1090 Vienna, Austria}
 \author{Aisling Johnson}
 \affiliation{University of Vienna,  Faculty of Physics, Vienna Center for Quantum Science and Technology, A-1090 Vienna, Austria}
 \author{Markus Aspelmeyer}
 \affiliation{University of Vienna,  Faculty of Physics, Vienna Center for Quantum Science and Technology, A-1090 Vienna, Austria}
 \affiliation{Institute for Quantum Optics and Quantum Information (IQOQI) Vienna, Austrian Academy of Sciences, A-1090 Vienna, Austria.}
 \author{Uro\v{s} Deli\'{c}}
 \email{uros.delic@univie.ac.at}
 \affiliation{University of Vienna,  Faculty of Physics, Vienna Center for Quantum Science and Technology, A-1090 Vienna, Austria}

\begin{abstract}

Light-matter interaction in the ultrastrong coupling regime can be used to generate exotic ground states with two-mode squeezing and may be of use for quantum enhanced sensing. Current demonstrations of ultrastrong coupling have been performed in fundamentally nonlinear systems. We report a cavity optomechanical system that operates in the linear coupling regime, reaching a maximum coupling of $g_x/\Omega_x=0.55\pm 0.02$. Such a system is inherently unstable, which may in the future enable strong mechanical squeezing. 

\end{abstract}
\maketitle

Cavity quantum electrodynamics in the weak coupling regime, where the light-matter coupling is significantly weaker than other characteristic energy rates in the system, is one of the most well-studied areas of quantum optics. Most interactions are described remarkably well by applying the rotating wave approximation that preserves only the number-conserving terms, resulting in the notable Jaynes-Cummings Hamiltonian \cite{JaynesCummings}. Once the coupling rate $g$ becomes comparable to the main transition frequency $\omega$ ($g\gtrsim \omega/10$), the quantum system enters the so-called ultrastrong coupling regime where the rotating wave approximation fails to fully describe the system dynamics \cite{forn-diaz_ultrastrong_2019,frisk_kockum_ultrastrong_2019}. Perhaps the most interesting consequence is that the ground states of ultrastrongly coupled systems become entangled as the counter-rotating terms that create pairwise excitations now contribute significantly to the system dynamics \cite{ciuti_quantum_2005,NatafGS,vitalireddetuned_2007,markovic_demonstration_2018}. In order to gain access to these exotic ground states, it is generally required to observe symmetry breaking of the quantum vacuum \cite{WangSymmetry} or to instantaneously tune the coupling rate \cite{VacantiVirtual,StassiVirtual,GarzianoVirtual}. This has been challenging to implement in experiments that have demonstrated operation in the ultrastrong coupling regime \cite{grossprb,YoshiharaDeepstrongQubit,VasanelliUSC,AskenaziUSC,BayerLandauUSC,MuellerPlasmonsUSC,BenzmoleculeOM,peterson_ultrastrong_2019,GeorgeMolecularMaterials,GambinoOrganicUCS,Singh_USC,DuanUSC}. As the coupling rate is further increased to above $g=0.5~\omega$, a quantum phase transition may occur \cite{forn-diaz_ultrastrong_2019,frisk_kockum_ultrastrong_2019}. At this point, the resulting nonclassical states of the quantum system become highly sensitive to external forces, which is why such systems have been proposed as a resource for critical metrology and quantum sensing \cite{garbeQST,pleniosensingprx,Gietka2022understanding,garbe_exponential_2022}. Intrinsically nonlinear systems will recover stability for coupling rates $g>0.5~\omega$ as the nonlinearity raises eigenfrequencies to real and positive values \cite{FelicettiNonlinearUSC,Bibak2022}. In stark contrast, a purely linear system will remain unstable for larger coupling rates as one normal mode effectively evolves in an inverted potential \cite{SudhirUnstableUSC}: the quantum state grows exponentially fast and becomes extremely susceptible to external disturbances \cite{WeissDelocalization,Cosco}. Exploiting inverted potential dynamics due to ultrastrong coupling has been out of reach of current experiments.

In this Letter, we demonstrate a purely linear system that operates in the ultrastrong coupling regime: an optically levitated nanoparticle coupled to an optical cavity via coherently scattered trapping light. We demonstrate full tunability of the coupling strength \textit{in situ}, paving the way for quantum sensing protocols and novel quantum control techniques in levitated optomechanics that exploit the linear nature of the interaction. 
 
Cavity optomechanics, where the cavity field interacts with the motion of a mechanical oscillator with a frequency $\Omega$, is a well-known paradigm of light-matter interaction \cite{optomechanicsRMP}. The standard cavity optomechanical interaction arises from the dispersive shift of the cavity resonance induced by the object motion. The shift of the cavity frequency effectively modifies the intensity of the driven cavity field, thus yielding a nonlinear optomechanical ("Kerr") Hamiltonian that couples the number of photons in the cavity and the oscillator. In the weak coupling limit ($g\ll \Omega$), linearizing the interaction results in the parametric coupling rate $g\propto \sqrt{n_{\text{phot}}}$ that is enhanced by the intracavity photon number $n_{\text{phot}}$ \cite{MarquardtCooling,WilsonRaeCooling,GenesCooling}. Optomechanical platforms have traditionally been designed to operate in the sideband resolved regime ($\kappa/4<\Omega$) as it allows for ground state cooling. In this case, by increasing the photon number one first achieves strong coupling ($g>\kappa/4$), which is required for coherent two-mode interaction \cite{Aspelmeyer_2009,TeufelStrong,VerhagenStrong,EnzianStrong,Quidant_2021}. Even stronger cavity drives are then required to reach ultrastrong parametric optomechanical coupling, where the linearized approximation breaks down \cite{Bibak2022}. Ultrastrong coupling has recently been realized in several dispersively coupled systems in the sideband resolved \cite{peterson_ultrastrong_2019,Singh_USC} and unresolved regime ($\kappa/4 >\Omega$) \cite{leijssen_nonlinear_2017,fogliano_mapping_2021,BenzmoleculeOM}.

Somewhat surprisingly, optomechanics also allows for genuinely linear interaction by using optically levitated dielectric nanoparticles or atoms, in a configuration where the polarizable object "coherently scatters" the trapping laser to drive the optical cavity from within \cite{Aspelmeyer_2019,Windey2019,GonzalezPRA}. In contrast to dispersive coupling, here the particle motion couples to the cavity electric field, therefore resulting in an inherently linear optomechanical Hamiltonian. This novel interaction mechanism has recently been used to cool a levitated nanoparticle to its motional quantum ground state in a room temperature environment \cite{Aspelmeyer_2020,Ranfagni2022,Piotrowski2023}. Furthermore, the possibility to simultaneously and independently tune the coupling rate and the mechanical frequency enables truly unique control techniques, which are of interest for thermodynamical studies \cite{DechantHeat,DechantMaxEff,landauer}, generation of strongly squeezed states beyond the 3 dB limit \cite{CernotikSqueezing} or matter-wave interference \cite{RomeroIsartInterference,Neumeier}. Being of fundamentally linear nature, the system Hamiltonian becomes unstable for $g> 0.5~\Omega$, which has recently been proposed in the context of strong mechanical squeezing \cite{kustura_mechanical_2022} that enables increased force sensitivity and entanglement \cite{WeissDelocalization,Cosco}.

\begin{figure}
    \centering
    \includegraphics[width=1\linewidth]{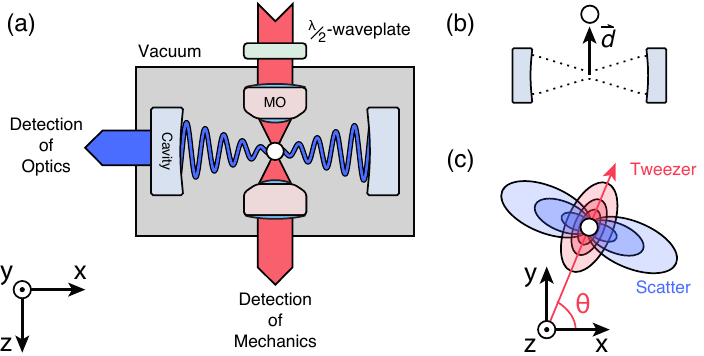}
    \caption{Experimental setup for cavity optomechanics via coherent scattering. (a) A silica nanoparticle is trapped in an optical tweezer (propagating along the $z$-axis, red) formed by a microscope objective (MO) inside the vacuum chamber. The particle scatters light into the empty cavity mode (oriented along the $x$-axis, blue) to generate the optomechanical interaction. The tweezer polarization is controlled by a $\lambda/2$-waveplate outside the chamber. The forward scattered light is recollimated and used in detection of the three orthogonal mechanical modes. The light leaking from the cavity is used to detect the optical mode. (b) The coupling rate can be changed by modifying the mean position of the nanoparticle $\vec{d}$ relative to the cavity waist with a 3-axis nanopositioner on which the microscope objective is mounted. (c) The angle of the tweezer polarization relative to the cavity mode polarization $\theta$ can be used to modify the amount of the scattered light -- and thus the coupling rate -- by rotating the aforementioned $\lambda/2$-waveplate.}
    \label{fig.apparatus}
\end{figure}

{\it Theory.}--We consider a dielectric nanoparticle with a polarizability $\alpha$ and a radius $r$ levitated in an optical tweezer given by the complex electric field amplitude $\vec{E}_\text{tw}$, and interacting with an optical cavity field $\vec{E}_\text{cav}$ (Figure \ref{fig.apparatus}). As long as $r\ll\lambda$, where $\lambda$ is the wavelength of the cavity mode of interest, one can simplify the optical interaction to the standard dipole interaction with the total electric field: $H_{\text{int}}=-\alpha|\vec{E}_\text{tw}+\vec{E}_\text{cav}|^2/2$ \cite{Aspelmeyer_2019,GonzalezPRA}. The interaction with the tweezer field $H_{\text{tw}}=-\alpha|\vec{E}_\text{tw}|^2/2$ is responsible for particle trapping, while the interaction with the cavity field $H_{\text{cav}}=-\alpha|\vec{E}_\text{tw}|^2/2$ leads to the standard dispersive coupling that is maximal at the slope of the cavity standing wave \cite{ChangProposal,RomeroIsartProposal,Aspelmeyer2_2020}. The interference term describes coherent coupling of the tweezer and cavity fields $H_{\text{coh}}=-\alpha\text{Re}(\vec{E}_\text{cav}\vec{E}^*_\text{tw})$, which yields a coupling that is maximal at the cavity node, i.e. where the cavity intensity is minimal. Restricting our attention to the relevant motional degree of freedom along the cavity ($x$-) axis and assuming that the particle motion is well-localized within a cavity standing wave period, the optomechanical Hamiltonian takes the linear form:
\begin{equation}
	\hat{H}/\hbar=\Omega_x\hat{b}^\dagger\hat{b}+\Delta\hat{a}^\dagger\hat{a}+g_x(\hat{a}+\hat{a}^\dagger)(\hat{b}+\hat{b}^\dagger).
	\label{Hamiltonian}
\end{equation}
Here, $\hbar$ is the reduced Planck constant, $\Delta=\omega_\text{cav}-\omega_\text{drive}$ is the detuning of the tweezer frequency $\omega_\text{drive}$ with respect to the cavity resonance $\omega_\text{cav}$, $\Omega_x$ is the bare frequency of the mechanical oscillator and $\hat{a}$ ($\hat{b}$) and $\hat{a}^\dagger$ ($\hat{b}^\dagger$) are the photon (phonon) annihilation and creation operators, respectively. The coupling rate $g_x(\vec{d},\theta)=E_d(\theta)kx_{\text{zpf}}f(\vec{d})$ is a function of the drive $E_d(\theta)=\alpha\sqrt{\frac{\omega_{\text{cav}}}{2\hbar\varepsilon_0 V_{\text{cav}}}}E_{\text{tw}}\cos (\theta)/2$, where $V_{\text{cav}}$ is the cavity mode volume, $\theta$ is the angle set by the polarization of the tweezer field relative to the cavity field, $x_{\text{zpf}}$ is the zero point fluctuation, $E_{\text{tw}}$ is the tweezer electric field and $f(\vec{d})$ is the cavity mode function at a position $\vec{d}$ relative to the central cavity node. In contrast to dispersive optomechanics, note that the coupling rate here is independent of the intracavity photon number, therefore it remains constant irrespective of the detuning. 

In our system the coupling rate can be tuned in three possible ways: (i) by changing the particle position $\vec{d}$ with respect to the cavity axis (Figure \ref{fig.apparatus}(b)), (ii) by turning the $\lambda/2$-waveplate outside the vacuum chamber to change the polarization angle $\theta$ and thus the amount of coherently scattered light into the cavity mode (Figure \ref{fig.apparatus}(c)) and (iii) by adjusting the tweezer power, which also modifies the mechanical frequency. Although we use only the first method in this work, we note that the other two methods in principle allow for instantaneous tuning of the coupling rate and have been demonstrated elsewhere \cite{Aspelmeyer_2019,Windey2019}. Furthermore, in dipole approximation the coupling rate depends on the particle radius as $g_x\propto r^{3/2}$, while the mechanical frequency to the leading order depends only on the particle density \cite{SI}. We will consider the case where the nanoparticle is positioned at the central cavity node while allowing for displacement $z_0$ along the tweezer axis with $\vec{d}=\begin{pmatrix}0,&0,&z_0\end{pmatrix}$, in which case the dispersive interaction is negligible while the coupling due to coherent scattering is maximized. The particle position at the cavity node can be further optimized by minimizing the cavity output power \cite{Aspelmeyer_2019}. 

For a red-detuned tweezer ($\Delta<0$) and weak coupling ($g_x\ll\Omega_x$) it is common to simplify the interaction to keep only the terms $\hat{a}\hat{b}^\dagger$ and $\hat{a}^\dagger\hat{b}$ that preserve the total energy as the counter-rotating terms contribute weakly to the Hamiltonian. However, as $g_x$ becomes comparable to $\Omega_x$, the rotating wave approximation is no longer valid as the perturbative approach fails. We recover the system dynamics in the Fourier domain from Eq. \eqref{Hamiltonian}:
\begin{equation}
\textbf{M}(\omega)\begin{pmatrix} \tilde{a}(\omega) \\ \tilde{a}^\dagger(\omega) \\ \tilde{b}(\omega) \\ \tilde{b}^\dagger(\omega) \end{pmatrix}=\begin{pmatrix} \sqrt{\kappa}\chi_l(\omega)\tilde{a}_\text{In}(\omega) \\ \sqrt{\kappa}\chi_l^*(-\omega)\tilde{a}_\text{In}^\dagger(\omega) \\ \sqrt{\gamma}\chi_m(\omega)\tilde{b}_\text{In}(\omega) \\ \sqrt{\gamma}\chi_m^*(-\omega)\tilde{b}_\text{In}^\dagger(\omega) \end{pmatrix},\label{eq.langevin}
\end{equation}
where $\textbf{M}(\omega)$ is the coupling matrix with
\begin{equation}
\textbf{M}(\omega)=\left(\begin{smallmatrix} 1 & 0 & ig\chi_l(\omega) & ig\chi_l(\omega) \\ 0 & 1 & -ig\chi_l^*(-\omega) & -ig\chi_l^*(-\omega) \\ ig\chi_m(\omega) & ig\chi_m(\omega) & 1 & 0 \\ -ig\chi_m^*(-\omega) & -ig\chi_m^*(-\omega) & 0 & 1 \end{smallmatrix}\right).
\end{equation}
Here, $\chi_l(\omega)=(\kappa/2-i(\omega-\Delta))^{-1}$ and $\chi_m(\omega)=(\gamma/2-i(\omega-\Omega_x))^{-1}$ are the complex optical and mechanical susceptibilities, respectively. $\kappa$ and $\gamma$ are the damping rates characterizing the coupling of the optical and mechanical subsystems to their photonic and phononic baths, modeled here as input fields $\hat{a}_\text{In}$ and $\hat{b}_\text{In}$, respectively. Note that generally the optical loss dominates over the mechanical loss ($\kappa\gg\gamma$). For a full solution see the Supplemental Information \cite{SI}. 

In the strong coupling regime ($g>\kappa/4$), the photonic and phononic modes hybridize, giving rise to so-called polariton modes with eigenfrequencies $\Omega_\pm$ and damping rates $\gamma_\pm$ that satisfy $|\textbf{M}(\Omega_\pm+i\gamma_\pm)|=0$. In a linear system a dynamical instability occurs when one normal mode acquires a negative spring constant (the eigenfrequency becomes imaginary), which is given by the following condition:
\begin{align}
    \left(\frac{g_x}{\Omega_x}\right)^2>\frac{1}{4}\left(\frac{\Delta}{\Omega_x}+\frac{\kappa^2}{4\Delta\Omega_x}\right).
    \label{stability}
\end{align}
In our system $\Omega_x\approx \kappa$, therefore we operate in the sideband resolved regime. In this case, reaching the strong coupling regime automatically implies that we operate in the ultrastrong coupling regime as well as $g_x>\kappa/4>0.1\Omega_x$. The stability condition from Eq.\eqref{stability} for the close-to-resonant detuning ($\Delta\sim \Omega_x$) is then simplified to $g_x>\sqrt{\Delta\Omega_x}/2$, which at the optimal detuning for cooling $\Delta\equiv \Omega_x$ yields $g_x/\Omega_x>0.5$ \cite{SudhirUnstableUSC}. 

\begin{figure}[h!]
    \centering
    \includegraphics[width=\linewidth]{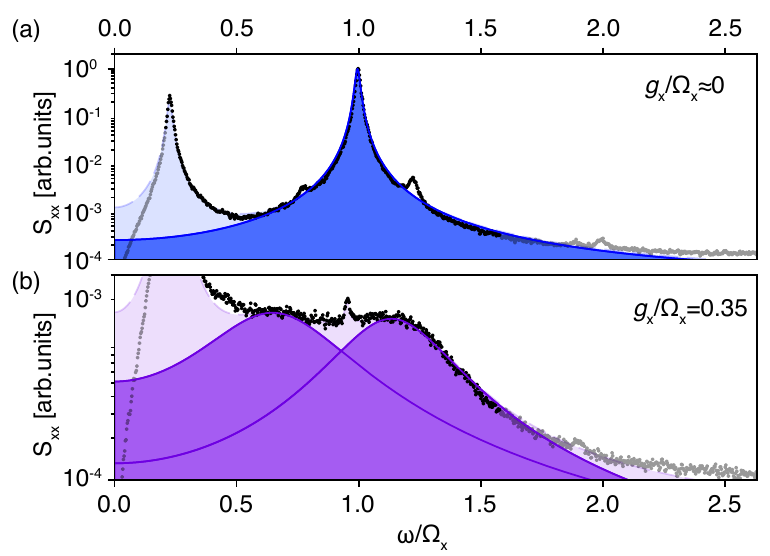}
    \caption{Power spectral densities (PSD) $S_{xx}(\omega)$ (arb. units) of the mechanical motion. (a) The standard PSD of a thermal mechanical oscillator that is unaffected by the cavity. Two peaks corresponding to the $z$- and $x$-motion are visible at $\Omega_z\approx0.2\:\Omega_x$ and $\Omega_x$, respectively. We fit the full PSD (light blue shaded) to extract the relevant PSD of the $x$-motion (dark blue shaded). (b) The mechanical mode exhibits normal mode splitting once the coupling is increased above $\kappa/4$. The full theoretical model (light purple shaded) is used to extract a coupling of $g_x/\Omega_x\approx 0.35$. The individual normal modes with frequencies $\Omega_\pm$ and dampings $\gamma_\pm$ are shown in dark purple. As the $x$-motion is significantly cooled, a narrow peak corresponding to the $y$-motion becomes visible at $\Omega_y\approx0.95\Omega_x$.}
    \label{fig.normal_mode_splitting}
\end{figure}

{\it Experiment.}---A schematic of our apparatus is shown in Figure \ref{fig.apparatus}. A high-finesse Fabry-P\'erot cavity (Finesse $\mathcal{F}=73000$, linewidth $\kappa=2\pi\times\SI{193}{k Hz}$, length $L=\SI{1.07}{cm}$, waist $w_\text{cav}=\SI{41.1}{\mu m}$) and a microscope objective (numerical aperture $\text{NA}=0.8$) mounted on a 3-axis nanopositioner (average step size $\sim 8~\text{nm}$) are mounted inside of a vacuum chamber \cite{Aspelmeyer2_2020}. The nanoparticle is trapped by an optical tweezer formed by focusing a laser beam (wavelength $\lambda=\SI{1064}{nm}$, optical power $P_\text{tw}\approx\SI{0.4}{W}$, waists $w_x=\SI{0.67}{\mu m}$ and $w_y=\SI{0.77}{\mu m}$) with the microscope objective, the focus of which is positioned close to the center of the cavity. A single nanoparticle with a nominal radius of $r=(105\pm 2)$ nm (microParticles GmbH) is trapped with mechanical frequencies of $(\Omega_x,\Omega_y,\Omega_z)=2\pi\times(190,180,40)~$kHz. In this work we operate at a pressure of $4~\text{mbar}$, where the mechanical dissipation $\gamma/2\pi\approx 4~\text{kHz}$ is dominated by collisions with the background gas.

{\it Normal Modes.}---We perform spectroscopy on the cavity and tweezer optical modes to reconstruct the normal modes. Light from the cavity output is mixed with a local oscillator to perform a heterodyne measurement, which allows us to extract the information about the hybridized optical mode. Furthermore, we recollimate the tweezer light after it has weakly interacted with the nanoparticle. This light can then be used in subsequent split detection to monitor all three mechanical modes \cite{SI}. We simultaneously record time traces of the photocurrents from all detections, which we use to calculate the power spectral density (PSD) of the hybridized modes. 

Figure \ref{fig.normal_mode_splitting} shows the appearance of normal mode splitting in the mechanical mode as we increase the coupling rate to $g_x>\kappa/4$. We first position the particle far away from the cavity mode where we don't observe coherently scattered light in the cavity output. For such weak coupling $g_x\ll \Omega_x,\kappa$ the cavity backaction is negligible, therefore the spectral feature of the particle motion in the PSD takes the form of a simple thermal harmonic oscillator (Figure \ref{fig.normal_mode_splitting}(a)). We then place the particle at a distance of $z_0\approx 0.6w_{\text{cav}}$ away from the cavity axis and set $\Delta=\Omega_x$, at which point we observe normal mode splitting (Figure \ref{fig.normal_mode_splitting}(b)). On its own, normal mode splitting clearly indicates that the system operates in the strong coupling regime. In the strong coupling approximation, the normal mode eigenfrequencies are symmetric around the intrinsic mechanical frequency $\Omega_x$. However, we observe that the normal modes are shifted toward lower frequencies, as predicted from the full solution to Eq.\eqref{eq.langevin} \cite{peterson_ultrastrong_2019}. We fit the spectra of the hybridized optical and mechanical modes to recover the normal mode frequencies \cite{SI}. We obtain the coupling of $g_x/\Omega_x=0.35$, which demonstrates that the system operates in the ultrastrong coupling regime.

\begin{figure}
    \centering
    \includegraphics[width=\linewidth]{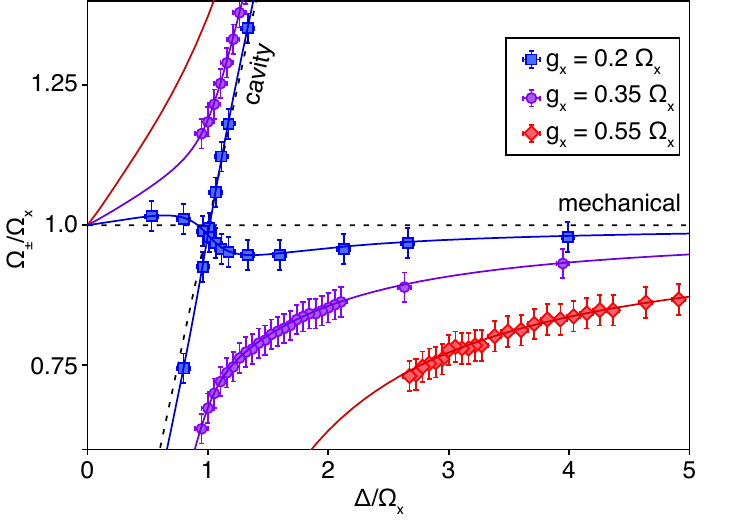}
    \caption{Measured normal mode frequencies (markers) and fits (solid lines) as a function of detuning. The decoupled cavity and mechanical mode frequencies are shown as the black dashed lines. We fit all markers from the same measurement set (markers with the same shape) with theory modelled with a single coupling rate \cite{SI}, which yields lines in strong agreement with our measurement. For a weakly coupled system ($g<\kappa/4$) the modes repel one another but still cross. This phenomenon still occurs at $g_x/\Omega_x \approx 0.2$ (blue). As the coupling increases to $g_x/\Omega_x\approx  0.35$ the system enters the strong coupling regime for $g\ge\kappa/4$ at which point the mechanical and optical modes hybridize, giving rise to new normal modes of the coupled system (purple). As the coupling is increased to $g_x/\Omega_x\approx 0.55$, the splitting grows and the lower eigenfrequency decreases significantly already for large detunings (red).}
    \label{fig.max_coupling}
\end{figure}

{\it Avoided crossing.}---We perform spectroscopy of the mechanical and optical modes as the detuning $\Delta$ is scanned towards the cavity resonance (Figure \ref{fig.max_coupling}). We jointly fit all PSDs in a detuning scan with a theory model with a single set of parameters to recover the coupling rate $g_x$ \cite{SI}. For weak coupling ($g_x=0.2~\Omega_x$) the modes repel each other ("optical spring"), but still cross as $g_x<\kappa/4$ (blue). As the detuning is tuned away from $\Delta/\Omega_x=1$, the normal mode frequencies converge to the bare frequencies of the optics and mechanics (black dashed lines), in agreement with theory. At a larger coupling rate ($g_x=0.35~\Omega_x$) the system reaches the strong coupling regime as $g_x>\kappa/4$. The modes strongly hybridize, which is confirmed by the avoided crossing of the normal mode frequencies (purple). At an even larger coupling rate of $g_x\approx 0.55~\Omega_x$ the avoided crossing is more prominent (red). Loss of the particle occurs at a detuning of $\Delta/\Omega_x\approx 2.6$, far away from the expected unstable region. We attribute this to small deviations of the particle position away from the cavity node that lead to a partial population of the cavity mode, which in turn pulls the particle further toward the cavity antinode. Experimental stability at lower detunings can be realized with a more careful placement of the particle at the cavity node. Although we were unable to probe the instability due to ultrastrong coupling, we note that it is only required to instantaneously jump to the unstable detuning and back to far detuning to use unstable dynamics for squeezing \cite{kustura_mechanical_2022}.

\begin{figure}
    \vspace{0.5cm}
    \centering
    \includegraphics[width=\linewidth]{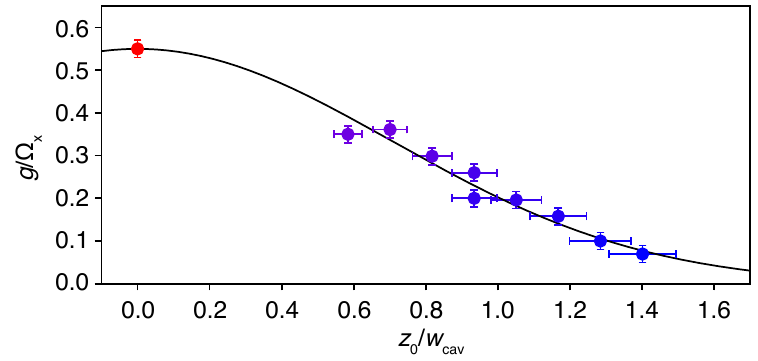}
    \caption{Spatial dependence of coupling via coherent scattering. As the nanoparticle is displaced away from the cavity axis, the overlap between the dipole scatter and the cavity mode should decrease following the cavity waist. Measured coupling rates (circles) decrease as the particle is displaced away from the cavity axis. The solid line is theory normalized to the highest observed coupling (red circle).}
    \label{fig.spatial_dependence}
\end{figure}

{\it Spatial dependence of coupling.}---We measure a maximum coupling of $g_x/\Omega_x= 0.55\pm 0.02$ for the particle placed directly on the cavity axis. Displacing the particle by $z_0$ away from the cavity axis maintains the form of the Hamiltonian to leading order, and simply reduces the coupling as $g_x(z_0)\propto \exp(-z_0^2/w_{\text{cav}}^2)$ \cite{SI}. We characterize the spatial dependence of the coupling by performing detuning scans as a function of $z_0$ (Figure \ref{fig.spatial_dependence}). The actual particle position $z_0$ has been previously calibrated through a measurement of the dispersive coupling \cite{Aspelmeyer2_2020}. We normalize the theoretical value of $g_x(z_0=0)$ to the maximum measured coupling, which yields a good agreement with the measured coupling over the entire spatial profile of the cavity mode.

{\it Conclusion and outlook.}---We have demonstrated linear ultrastrong coupling between the motion of a solid-state object and an optical cavity field in the resolved sideband regime. We have characterized the high degree of \textit{in situ} tuning of the system parameters, which we use to explore our system in all relevant regimes of coupling: from weak over strong to ultrastrong coupling regime. We are able to recover the full system dynamics of the optical and mechanical modes by combining several detection schemes. This allows us to obtain a maximum coupling of $g_x/\Omega_x=0.55\pm 0.02$ for a particle with a radius of $105$ nm. At this coupling rate, the system exhibits a dynamical instability when $\Delta\approx\Omega_x$, a necessary requirement for quantum sensing protocols \cite{garbe_exponential_2022}. For coupling slightly below $g_x/\Omega_x=0.5$ we expect ground state cooling and optomechanical entanglement to exist simultaneously \cite{vitalireddetuned_2007,markovic_demonstration_2018}. This will in future allow for detailed studies of global master equations and thermodynamics in presence of ultrastrong coupling \cite{KonopikPRR,Pilar2020thermodynamicsof}. 

There exist multiple avenues to reach even larger coupling $g_x/\Omega_x$. Reducing the tweezer power modifies both the coupling rate and the mechanical frequency, such that their ratio increases as $g_x/\Omega_x\propto P_\text{tw}^{-1/4}$. Alternatively, trapping larger particles would increase the coupling rate. Conservative estimates using the same system parameters -- taking radiation pressure and Mie scattering into account -- promise deep strong coupling rates ($g_x/\Omega_x>1$) for $r\geq \SI{150}{nm}$ \cite{SI}. Finally, using cavities with smaller mode volumes will increase the coupling by several orders of magnitude to $g_x/\Omega_x=10^2$, which would result in strong mechanical squeezing by $30$ dB due to unstable dynamics \cite{kustura_mechanical_2022}.

{\it Acknowledgments.}---We thank John Teufel, Peter Rabl, Louis Garbe, Fatemeh Bibak and Anton Zasedatelev for valuable discussions. This project was supported by the European Research Council (ERC CoG QLev4G and ERC Synergy Q-Xtreme), by the ERA-NET programme QuantERA, QuaSeRT (Project No. 11299191; via the EC, the Austrian ministries BMDW and BMBWF and research promotion agency FFG), by the Austrian Science Fund (FWF) and the doctoral school CoQuS (Project W1210) and the ESQ Discovery Grant "Ultrastrong cavity optomechanics" of the Austrian Academy of Sciences. 


%

\onecolumngrid
\appendix
\newpage
\makeatletter
\renewcommand*{\@biblabel}[1]{\hfill#1.}
\makeatother

\setcounter{figure}{0}
\setcounter{equation}{0}
\renewcommand{\thefigure}{S\arabic{figure}}
\renewcommand{\theequation}{S\arabic{equation}} 

\section{\Large Supplementary information}

\subsection{Experimental setup}

In this section we present our experimental setup in more details.

\begin{figure}[!h]
    \centering
    \includegraphics[width=.7\linewidth]{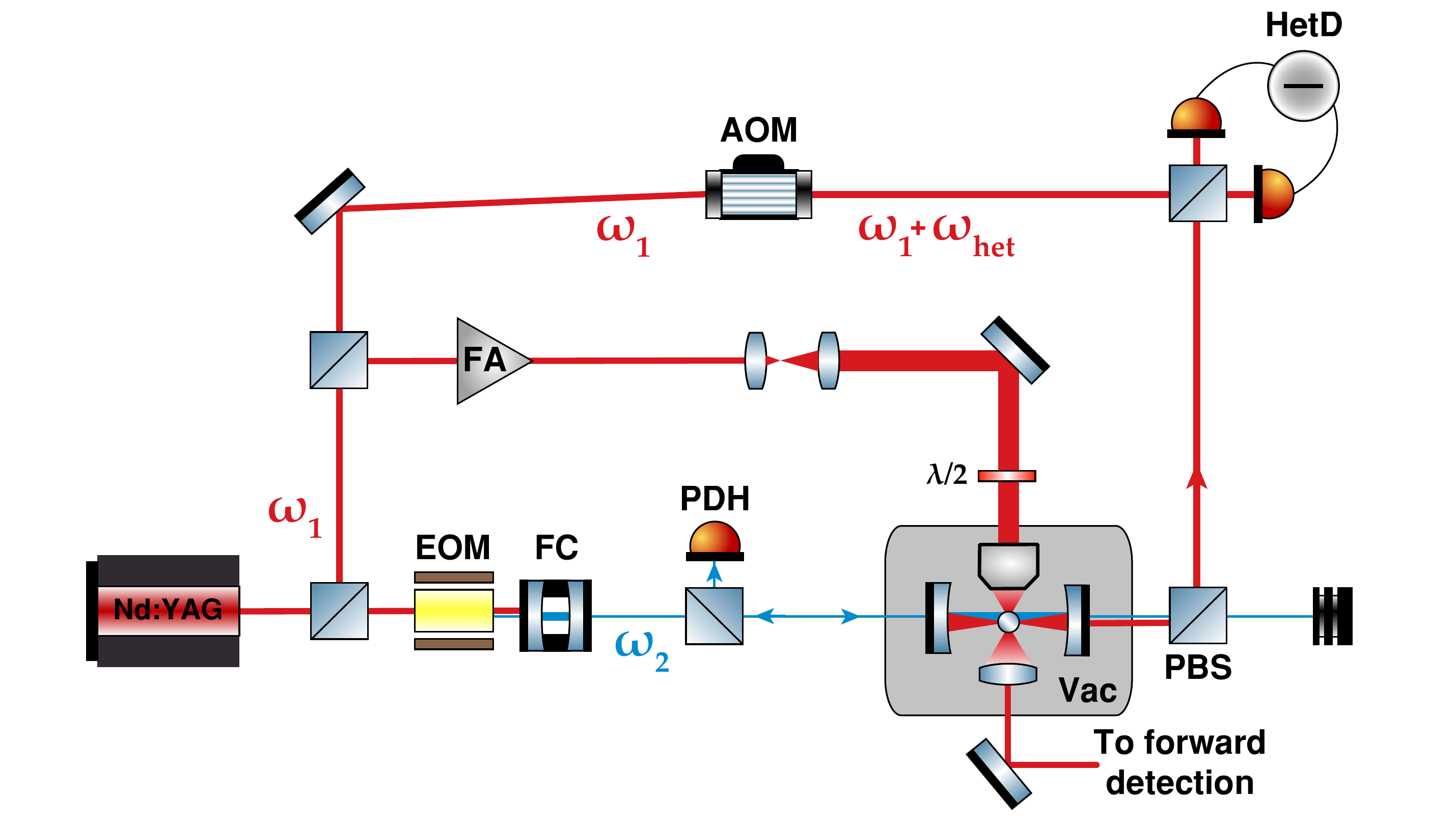}
    \caption{An Nd:Yag laser with a power output of \SI{2}{W} with a wavelength of \SI{1064}{nm} is split in three paths. The optical tweezer path contains a fiber amplifier (FA). The tweezer is formed by tightly focusing 400 mW of optical power with a high numerical aperture (NA = 0.8) microscope objective (MO) with overfilled entrance aperture. The polarization of the tweezer is controlled by a half-wave plate ($\lambda/2$) in front of the vacuum chamber (vac). A split detection scheme in forward direction is used to extract information about mechanical motion in all three directions. In the locking path, an electro-optical modulator (EOM) and a filtering cavity (FC) are used to create a sideband with frequency $\omega_2$, which is then used to stabilize the cavity length using a Pound-Drever-Hall (PDH) method. In heterodyne detection (HetD) path an acousto optical modulator (AOM) is used to generate a local oscillator to detect the cavity mode split at the polarizing beam splitter (PBS). Note that the setup is very similar to the one in \cite{Uros_Groundstate}.}
    \label{fig_setup}
\end{figure}

The light source is a \textbf{ND:YAG} (Innolight Mephisto) laser with an output power of \SI{2}{W} at a wavelength of \SI{1064}{nm} and frequency $\omega_1$. The laser power is split into three main paths to operate the experiment as illustrated in Figure \ref{fig_setup}: cavity locking, heterodyne detection, and the tweezer paths.

The tweezer power is increased by a high--power fiber amplifier (Keopsys, maximum output power 10 W). The optical tweezer is formed by tightly focusing 400 mW of optical power with a high numerical aperture (NA = 0.8) microscope objective (MO) with overfilled entrance aperture. This forms a three-dimensional trap with effective sizes of $(w_x, w_y, w_z) = (0.67, 0.77, 1.7)\,$\SI{}{\micro m}. The MO is mounted on a three-dimensional piezo translation stage (Mechonics MX35) with a travel range of \SI{25}{mm} and a step size of \SI{\sim8}{nm}, which allows for precise positioning of the nanoparticle within the cavity mode. The tweezer is linearly polarized with its polarization axis controlled by a half--wave plate placed before the MO.

For coherent scattering experiments, it is crucial to control the detuning between the tweezer light and the cavity resonance. It is also important to separate light used for locking the cavity and light scattered into the cavity mode by the nanoparticle. To meet both requirements, the laser is modulated with an electro-optical modulator (EOM) to create sidebands with frequencies $\omega = \omega_1 \pm \Delta \omega$, where $\Delta \omega = 2 \pi \cdot 14.0192 \, \text{GHz}$ is the free spectral range (FSR) of the optomechanical (OM) cavity. The upper sideband is then filtered by a filtering cavity (FC) and used to lock the laser light to the resonance of the OM cavity using the Pound-Drever-Hall (PDH) technique \cite{pdh_stabilization_black}. The polarization of the locking mode is orthogonal to the polarization of the coherent scattering mode.

We use a split detection scheme in the forward direction to extract information about the mechanical motion in all three directions and access the lower polariton branch. The tweezer light is recollimated with a low-NA ($\text{NA} = 0.2$) lens. A fraction of the tweezer beam interacts with the particle and carries information about its motion, the rest does not interact with it and serves as the local oscillator. The particle motion along the tweezer axis is detected directly by focusing the light on a photodiode. The motion in the transverse directions is detected with a split detection scheme where the recollimated tweezer light is split using a D-shaped mirror and then focused on a balanced photodetector (Thorlabs, PDB420C-AC). A Dove prism is used to rotate the recollimated beam to minimize the cross-talk between the two transverse motions.

The upper polariton for large detunings is very weakly imprinted on the mechanical motion but can be accessed by heterodyne detection of the cavity mode. An acousto-optical modulator (AOM) is used to shift the laser frequency $\omega_1$ by $\omega_{\textnormal{het}}=2\pi \cdot \SI{10}{MHz}$ in order to create a local oscillator (LO).

The experiment is run at a pressure of $\SI{4}{mbar}$. At this pressure the particle moves as an underdamped harmonic oscillator as $\Omega / \gamma \approx 500$. Furthermore, the cavity remains the dominant loss channel, as $\kappa / \gamma \approx 500$.


\subsection{Theory}
\begin{mdframed}
\begin{center}
\begingroup
\setlength{\tabcolsep}{10pt} 
\renewcommand{\arraystretch}{1.5} 
\begin{tabular}{ r@\quad r@\quad l@\quad}
Zero point fluctuation & $x_{\textnormal{zpf}}=$&$\sqrt{\frac{\hbar}{2m\Omega_x}}$ \\ 
Wave number & $k =$&$ 2 \pi/ \lambda$  \\  
Rayleigh range & $z_{\textnormal{R}} =$&$ w_x w_y \pi / \lambda$ \\
Cavity drive & $E_{\textnormal{d}}(\theta)=$&$ \alpha \epsilon_{\textnormal{tw}} \epsilon_{\textnormal{cav}} \sin \theta/(2\hbar)$\\
Cavity field &  $\epsilon_{\textnormal{cav}}=$&$ \sqrt{\frac{\hbar \omega_{\textnormal{cav}}}{2 \epsilon_0 V_{\textnormal{cav}}}}$\\
Tweezer field & $\epsilon_{\textnormal{tw}}=$&$ \sqrt{\frac{4P_{\textnormal{tw}}}{w_x w_y \pi \epsilon_0 c}}$ 
\end{tabular}
\endgroup
\end{center}
\end{mdframed}
The interaction of two coupled oscillators is described by the following linearized interaction Hamiltonian in terms of the creation and annihilation operators of the light field $\hat{a},\hat{a}^\dagger$ (\cite{theory_coherent_scattering}, \cite{Gonzalez_Ballestro_PRA}):
\begin{align}
    \hat{H}_{\textnormal{int}} &= E_{\textnormal{d}}(\theta)(\hat{a}^\dagger e^{i(kz-\phi G(z))}+\hat{a}e^{-i(kz-\phi G(z))})\cos{k(x_0+x\sin{\theta}+y\cos{\theta})}\\
    &=-(\hat{a}^\dagger + \hat{a})\left [\underbrace{(E_{\textnormal{d}}(\theta)kx_{\textnormal{zpf}}\sin{\theta}\sin{k x_0}}_{g_x(\theta,x_0)}\frac{\hat{x}}{x_{\textnormal{zpf}}}f(\vec{x}_0)+\underbrace{(E_{\textnormal{d}}(\theta)ky_{\textnormal{zpf}}\cos{\theta}\sin{k x_0}}_{g_y(\theta,x_0)}\frac{\hat{y}}{y_{\textnormal{zpf}}}f(\vec{x}_0)\right]\\
    &+i(\hat{a}^\dagger -\hat{a})\underbrace{E_{\textnormal{d}}(\theta)(k-1/z_{\textnormal{R}})z_{\textnormal{zpf}}\cos k x_0}_{g_{z}(\theta,x_0)} \frac{\hat{z}}{z_{\textnormal{zpf}}}f(\vec{x}_0)+\underbrace{E_{\textnormal{d}}(\theta)\cos kx_0 (\hat{a}^\dagger + \hat{a})}_{\text{dispersive interaction}}.
    \label{eq:all_coupling}
\end{align} 
Here, $\hat{x}$ denotes the nanoparticle position operator along the x-axis. Note that this Hamiltonian is identical to \cite{theory_coherent_scattering}, but with an additional geometrical factor 
\begin{align}
     f(\vec{d})&=\frac{e^{-\frac{y_0^2 + z_0^2}{w_\text{cav}^2}}}{\sqrt{\frac{W_{x}(z_\text{rad})}{w_x}\frac{W_{y}(z_\text{rad})}{w_y}}}.
     \label{eq:geom_factor}
\end{align}
Here $W_x(z_{\textnormal{rad}})$ ($W_y(z_{\textnormal{rad}})$) denotes trap waist in the x (y) direction at particle equilibrium position along the trap axis $z_{\textnormal{rad}}$ due to the radiation pressure. This geometric factor is used to describe how far the current equilibrium position is located from the tweezer and cavity waists, such that at the trap and cavity center per definition $\vec{d}=0$ and $f(0) = 1$. 
At the mean particle position $\vec{d}$ the coupling strength in the x-direction is given by:
\begin{align}
     g_x(\vec{d})&=kx_\text{zpf}\sqrt{\frac{2}{\pi}\frac{P_\text{tw}}{\hbar }\frac{k}{w_xw_y}\frac{\alpha}{\epsilon_0}\frac{4}{w_\text{cav}^2\pi L}\frac{\alpha}{\epsilon_0}}f(\vec{d}).
     \label{eq:coupling_x}
\end{align}
Here, $P_\text{tw}$ represents the optical tweezer power, $w_\text{x} \, (w_\text{y})$ is the waist of the optical tweezer along the x-(y-) axis, $w_\text{cav}$ is the waist of the cavity mode, $\epsilon_0$ is the vacuum permittivity, and $L$ is the cavity length. 

In our experiments the particle is positioned at the cavity node for which per definition $\cos{kx_0}=0$ and the z-coupling and dispersive interaction are minimized. The tweezer polarization is selected to be orthogonal to the cavity axis, such that $\cos{\theta}=0$. Therefore, only the nanoparticle x-motion (along the cavity axis) is coupled to the cavity mode,

\begin{align}
    \hat{H}_{\textnormal{int}}= \hbar g_x (\hat{a} + \hat{a}^\dagger)(\hat{b} + \hat{b}^\dagger).
\end{align}
Here $\hat{b},\,\hat{b^\dagger}$ are the phonon annihilation and creation operators, such that $(\hat{b} + \hat{b}^\dagger)=\hat{x}/x_{\text{zpf}}$. The counter-rotating terms $\hat{a}\hat{b}$ and $\hat{a}^\dagger \hat{b}^\dagger$ cannot be neglected in the strong and ultrastrong coupling regimes. The equations of motion are extracted from the full Hamiltonian $H_{\textnormal{cav}}+ H_{\textnormal{mech}} + H_{\textnormal{int}}$ using the approach described in \cite{Caldeira_Leggett}, 
\begin{align}
    \Dot{\hat{a}}&\approx - \left(\frac{\kappa}{2}+i\Delta'\right) \hat{a} + \sqrt{\kappa}\hat{a}_{\textnormal{in}} - ig (\hat{b}+\hat{b}^\dagger)\\
    \Dot{\hat{b}}&\approx - \left(\frac{\gamma}{2}+i\Omega_x'\right)\hat{b} + \sqrt{\gamma}\hat{b}_{\textnormal{in}} - ig (\hat{a}+\hat{a}^\dagger).                     
\end{align}
We define optical and mechanical susceptibilities
\begin{align}
    \chi_{\text{l}} &= \frac{1}{\frac{\kappa}{2}-i(\omega-\Delta)}\\
    \chi_{\text{m}} &= \frac{1}{\frac{\gamma}{2}-i(\omega-\Omega)},
\end{align}
and in the Fourier domain the coupled equations of motion can be written in a  matrix form:
\begin{align} \label{eq:matrixform}
\left( \begin{array}{c}
\sqrt{\kappa}\chi_{\text{l}}(\omega) \Tilde{a}_{\text{in}}(\omega)\\ 
\sqrt{\kappa}\chi_{\text{l}}^*(-\omega) \Tilde{a}_{\text{in}}^\dagger(\omega)\\
\sqrt{\gamma}\chi_{\text{m}}(\omega) \Tilde{b}_{\text{in}}(\omega) \\ 
\sqrt{\gamma}\chi_{\text{m}}^*(-\omega) \Tilde{b}_{\text{in}}^\dagger(\omega)
\end{array}\right)=\underbrace{
\left( \begin{array}{cccc}
1 & 0 & ig\chi_{\text{l}}(\omega) & ig\chi_{\text{l}}(\omega)\\
0 & 1 & -ig\chi_{\text{l}}^*(\omega) & -ig\chi_{\text{l}}^*(-\omega)\\
ig\chi_{\text{m}}(\omega) & ig\chi_{\text{m}}(\omega) & 1 & 0 \\
-ig\chi^*_{\text{m}}(-\omega) & -ig\chi^*_{\text{m}}(-\omega) & 0 & 1
\end{array}\right) }_{M(\omega)}
\left( \begin{array}{c}
\Tilde{a}(\omega)\\
\Tilde{a}^\dagger(\omega)\\
\Tilde{b}(\omega)\\
\Tilde{b}^\dagger(\omega)
\end{array}\right).
\end{align}
In the ultrastrong coupling regime, the optical and mechanical modes strongly hybridize into an upper and lwer polariton branches. The frequencies of these two modes can be computed by solving
\begin{align} \label{eq:det_eqn}
    |M(\Omega_\pm + i \gamma_\pm)| = 0,
\end{align}
where $\Omega_\pm$ are the two eigenfrequencies of the upper and lower polaritons, respectively, and $\gamma_\pm$ are the corresponding damping rates. The equation can be rewritten as
\begin{align}
    0= \left(\frac{\lambda^2-\Omega^2+i\Gamma\lambda}{2\Omega}\right)\left(\lambda-\Delta+\frac{i\kappa}{2}\right)\left(\lambda+\Delta+\frac{i\kappa}{2}\right)+2\Delta g^2.
\end{align} 
The power spectral density (PSD) of the mechanical oscillator is then deduced from Equation \ref{eq:matrixform}: 
\begin{align} \label{eq:psd_x}
    S_{xx}(\omega)= x^2_{\text{zpf}}|\nu(\omega)|^2 \left[\gamma|\chi_{\text{m}}(\omega)|^2(\Bar{N}+1)+\gamma|\chi_{\text{m}}(-\omega)|^2\Bar{N} + g^2\kappa |\chi_{\text{l}}(\omega)|^2|\chi_{\text{m}}(\omega)-\chi_{\text{m}}^*(-\omega)|^2\right],
\end{align}
where 
\begin{align}
    \nu(\omega) = \frac{1}{1 + g^2 \left(\chi_{\text{l}}(\omega)-\chi_{\text{l}}^*(-\omega)\right)\left(\chi_{\text{m}}(\omega)-\chi_{\text{m}}^*(-\omega)\right)}.
\end{align}

Equation \ref{eq:psd_x} is derived using the full USC theory. It is used to fit the experimentally measured PSDs and deduce the polariton frequencies $\Omega_\pm$. Equation \ref{eq:det_eqn} provides an exact theoretical expression for the polariton frequencies, which we then fit to the measured frequencies to obtain all setup parameters.

\subsection{Scaling of the coupling strength with particle size}

One of the possible avenues to realize larger coupling strengths is to increase the particle radius $r$. The  coupling rate is proportional to particle polarizability and the zero point fluctuation, which in total scales with $r^{3/2}$. However, one cannot increase the particle size indefinitely, since at some point the dipole approximation breaks and the coherent scattering into the cavity mode deviates from an ideal dipole radiation. Moreover, larger particles experience a stronger radiation pressure force such that its trapping position is further away from the tweezer waist. As the intensity at the trapping position decreases with the distance from the waist, this also decreases the coupling rate and in addition makes trapping more challenging. In the following, we perform numerical simulations to estimate of the scaling of the coupling rate with the particle size.

\begin{figure} [ht!]
        \centering
        \includegraphics[width=0.5\linewidth]{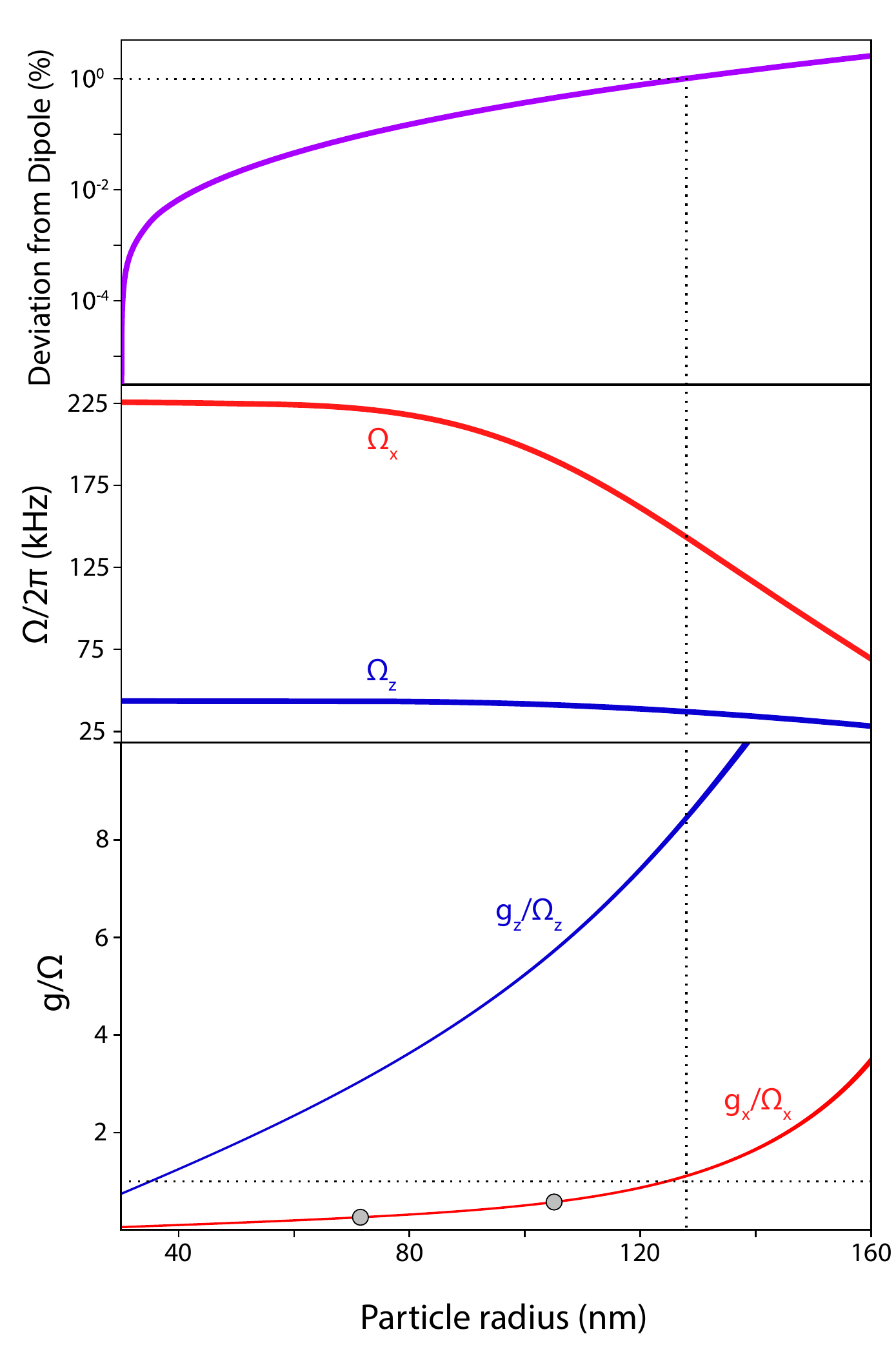}
        \caption{Modeling of how the coherent scattering coupling scales with particle size. We compute the deviation of the Mie solution from an ideal dipole (top). We see good agreement between the two solutions even up to particles radii of $r\approx\SI{160}{nm}$. We also model the effect of radiation pressure on the equilibrium trap frequencies for the x- and z-modes (middle). Both of these effects give estimates for the scaling of the coupling with particle size (bottom). The two circles here represent values measured in our experiment with particles of radii $r=71.5\,\text{nm}$ and $r=105\,\text{nm}$. The x-mode should enter the DSC regime for particles of radius $r\approx\SI{127}{nm}$ where the Mie solution deviates from the dipole approximation by a few percent. The effect of this deviation is given by the shaded region of the estimates, as a full calculation of the higher order couplings would be required to more precisely predict the scaling.}
        \label{fig:coupling_scaling}
    \end{figure}

To do so we simulate the dipole trap using the Optical Tweezers Toolbox \cite{Nieminen_2007}. First, numerical aperture, trapping power and beam truncation angle are set by simulating the trapping frequencies and matching them to the experimentally measured values in \cite{Uros_Groundstate}. The following system parameters were found to produce the best match to the experimental results: $P_{\text{tw}}=\SI{500}{mW},\, NA_{\textnormal{tw}}=0.60,\, NA_{\textnormal{beam}}=0.65$. The mismatch between the tweezer and beam numerical apertures comes from the overfilled objective.

To estimate the validity of the dipole approximation, the following simulation is performed. We simulate the field scattered by a particle of  \SI{30}{nm} radius fixed at the trap waist. Then, the power scattered in the direction orthogonal to both the trap axis and trap polarization is measured, and normalized by the total power scattered by the particle. This direction is chosen as it is the cavity axis in the experiment. Since a \SI{30}{nm} particle is well within the dipole approximation, we consider it an ideal dipole. The simulation is then repeated for the particle radii varying from 70 to \SI{160}{nm}, and the scattered power is compared to the ideal dipole case. 

The coupling rate is estimated using Equations \ref{eq:geom_factor} and \ref{eq:coupling_x}. The trap waist at the trapping position is estimated from the simulation, as well as the particle frequencies. The result is represented in Figure \ref{fig:coupling_scaling}.

The simulation suggests that \SI{99}{\%} of the light scattered by a $\SI{127}{nm}$ radius particle is in the dipole mode (top graph). Coupling strength $g_x/\Omega_x$ predicted by the simulation (bottom graph, red) is cross-referenced with the measurements performed in our experiment (bottom graph, gray circles). To do this, we set $g_x/\Omega_x = 0.265$ for $r=71.5\,\text{nm}$ particles, which is a value measured in our experiment. The simulation then predicts $g_x/\Omega_x=0.59$ for $r=105\,\text{nm}$ particles, while the measured value is $g_x/\Omega_x=0.55$. For a $\SI{127}{nm}$ radius particle we predict a coupling rate $g_x = 1.1\:\Omega_x$, which places our system into the deep strong coupling regime. In conclusion, it should be possible to reach even larger coupling strengths by trapping larger particles.

\section{Determination of the coupling strength}
In order to measure the coupling rate, the detuning is scanned from 5 MHz towards $\approx\Omega_x$, because for smaller detunings the system becomes unstable. At each detuning the time trace of the particle motion is recorded for all three directions using the forward detection scheme. Simultaneously, the light transmitted through a cavity mirror is detected using a heterodyne measurement. The bare mechanical frequency of an uncoupled oscillator is taken as $\Omega_x$ for $\Delta=\SI{5}{MHz}$. All the acquired spectra are fit simultaneously using Equations \ref{eq:det_eqn} and \ref{eq:psd_x} with the coupling strength as the only free parameter. This allows us to extract the coupling rate $g_x$.

\begin{figure}[h]
    \centering
    \includegraphics[width=.6\linewidth]{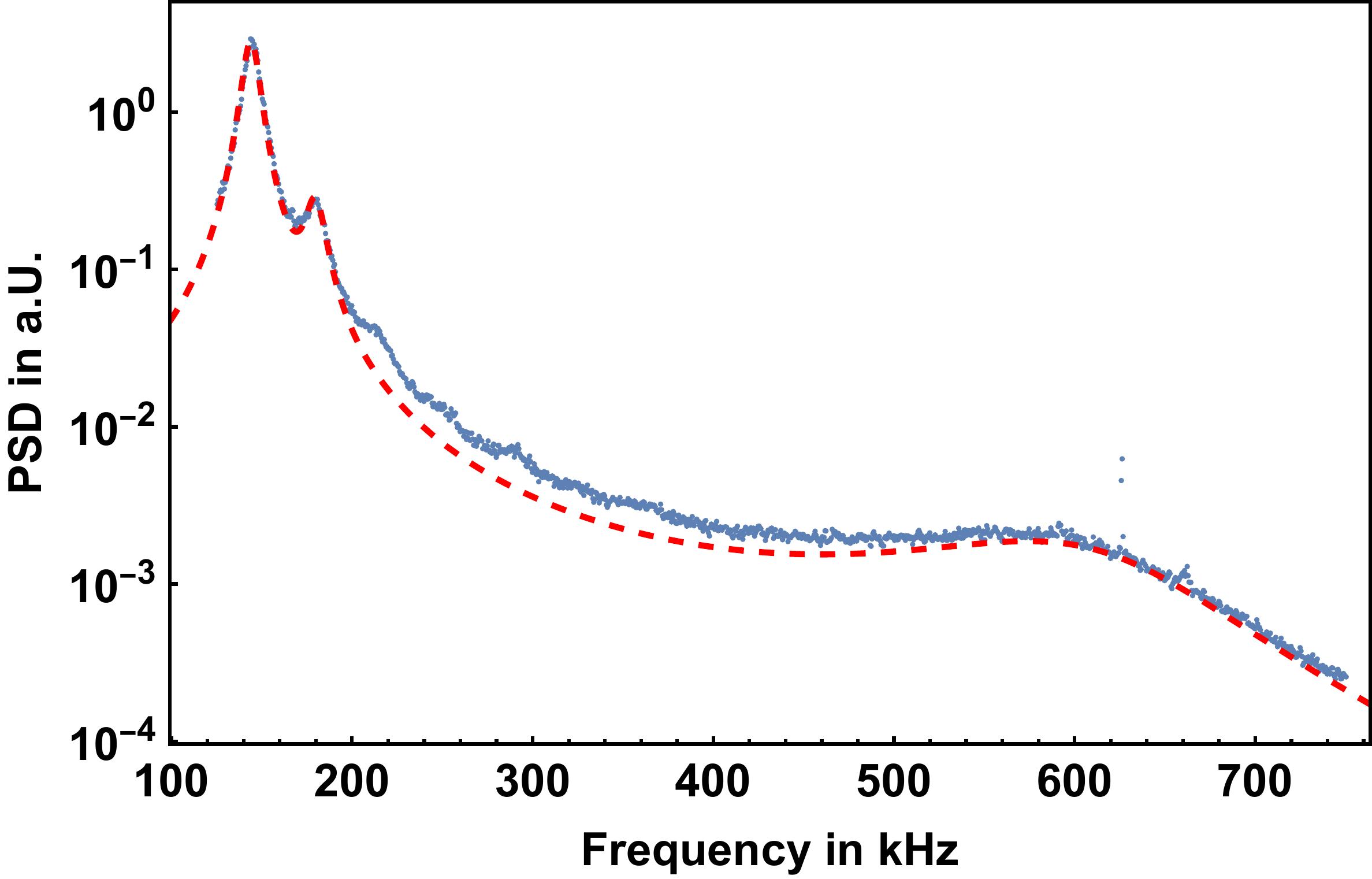}
    \caption{An example of a spectrum obtained with the heterodyne measurement. The x-axis shows the Fourier frequency in kHz, while the y-axis represents power the spectral density in arbitrary units. The blue line depicts the measurement, and the red dashed line the fit. One can observe peaks  x-motion (the y-motion is not coupled to the cavity), as well as its mixing with the z frequency. The cavity envelope is visible around \SI{600}{kHz}, corresponding to the detuning.}
    \label{fig_het}
\end{figure}

The detuning corresponding to each time trace is set manually, but can be measured using the heterodyne detection (Figure \ref{fig_het}). Here, the detuning is $\Delta/2\pi \approx \SI{600}{kHz}$, where the cavity transmission peak is visible. Because of the imperfections of the cavity lock, the cavity resonance drifts and therefore the detuning drifts are a major contribution to the measurement error.

\begin{figure}[!h]
    \centering
    \includegraphics[width=.45\linewidth]{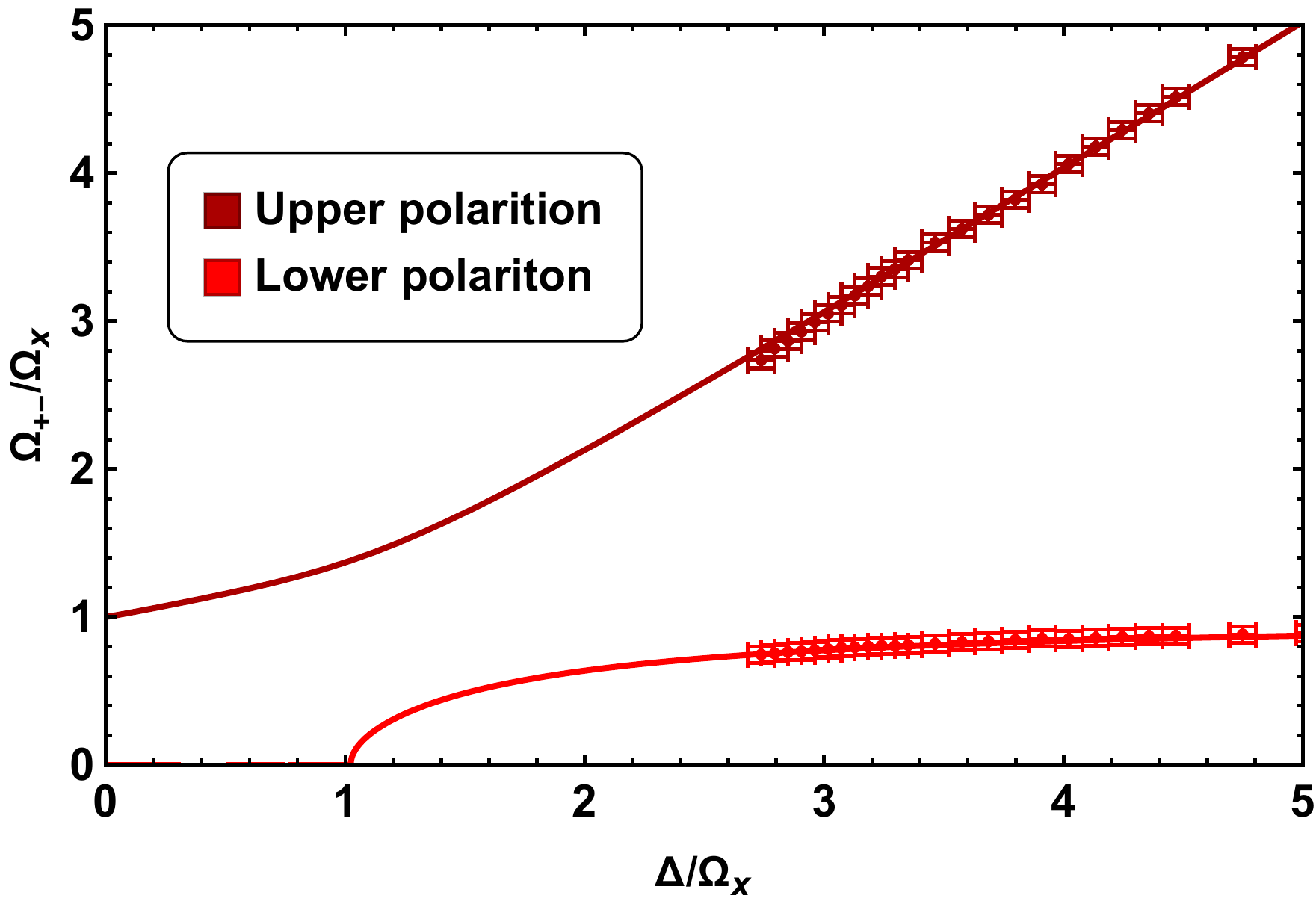}
    \caption{The upper polarition and the lower polariton of the coupled system with the coupling-strength of $g=0.55 \Omega_x$ are fit in the heterodyne spectrum for different detuning. The solid line represent the theory line for the given system while the data is plotted on top with error-bars which are mostly represented by the uncertainty in the detuning frequency.}
    \label{fig:upper_lower}
\end{figure}

For the strongest coupling ($g=0.55 \Omega_x$) we could not tune in far enough to be able to see the upper polariton in the forward detection because it mostly consists of the cavity mode. In the Heterodyne detection on the other hand one can observe the upper polariton as shown in figure \ref{fig_het}. In figure \ref{fig:upper_lower} we show the frequencies of the upper and lower polariton dependent on the detuning. 

\begin{figure}[!h]
    \centering
    \includegraphics[width=.5\linewidth]{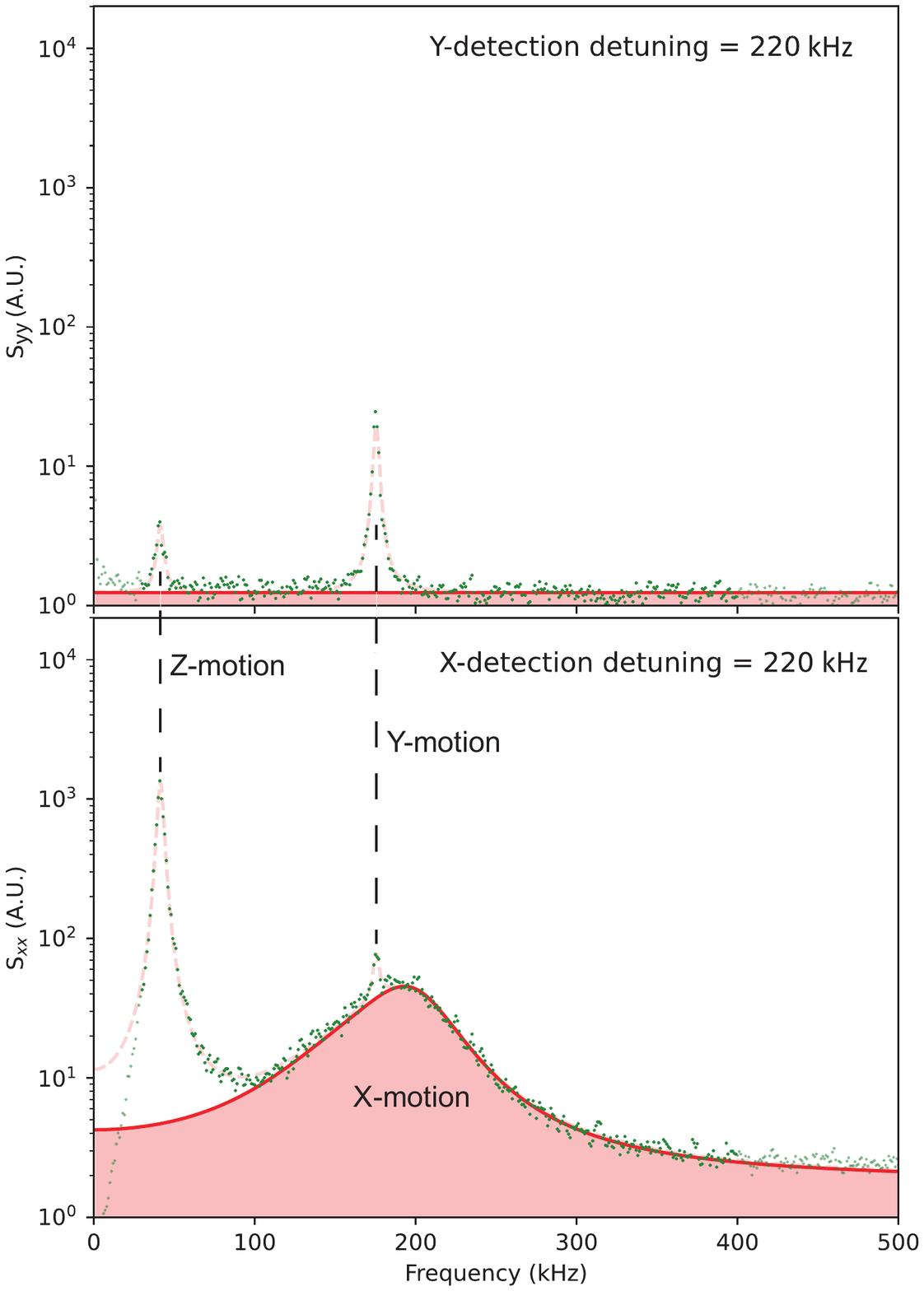}
    \caption{The spectra obtained in forward detection. The Fourier frequency is represented on the  horizontal axis, and the PSD (in arbitrary units) on the vertical axes. The data is shown in dotted-green and the fit in red. In the upper plot, the detection along the y-axis is shown, where the x-motion is well suppressed. Using this, the frequency and width of the z and y-peaks are extracted independently and used to fit the full spectra in the x-detection (bottom plot). The y-peak here is barely visible because of the dominant broadened x-peak.}
    \label{fig:independent_detection}
\end{figure}

As mentioned above, the forward detection is optimized to minimize cross-talk between different directions of motion. In Figure \ref{fig:independent_detection} (top) one can observe the spectrum obtained with the y-detector. It is evident that the broad x-peak is not visible here. However, this is not the case for the y-peak in x-detection (Figure \ref{fig:independent_detection}, bottom). Because the y-motion is not coupled to the cavity mode, it is not dampened and the sharp peak of the y-motion is always visible in the x-detection. To improve the fit quality of the x-peak, the central frequency and the width of the y- and z-peaks are extracted from the y-detection. The x-detection is then modeled as a sum of three peaks: two Lorentzians corresponding to z-motion and y-motion, and one modeled using full USC theory, corresponding to the x-motion.

\clearpage

\end{document}